\documentclass[%
 notitlepage,
 reprint,
 superscriptaddress,
 amsmath,amssymb,
 aps,
]{revtex4-2}

\usepackage{lipsum}
\usepackage{graphicx}
\usepackage{dcolumn}
\usepackage{bm}

\usepackage{epstopdf}
\usepackage{xcolor}

\newcommand{\bra}[1]{\langle #1 \vert}
\newcommand{\ket}[1]{\vert #1 \rangle}

\begin{document}

\preprint{APS/123-QED}

\title{Non-Rayleigh signal of interacting quantum particles} 

\author{M. F. V. Oliveira}
\affiliation{%
 Instituto de F\'{i}sica, Universidade Federal de Alagoas, 57072-900 Macei\'{o}, AL, Brazil
}%

\author{F. A. B. F. de Moura}
\affiliation{%
 Instituto de F\'{i}sica, Universidade Federal de Alagoas, 57072-900 Macei\'{o}, AL, Brazil
}%

\author{A. M. C. Souza}
\affiliation{%
 Departamento de F\'{i}sica, Universidade Federal de Sergipe, 49100-000 S\~{a}o Crist\'{o}v\~{a}o, SE, Brazil
}%

\author{M. L. Lyra}
\affiliation{%
 Instituto de F\'{i}sica, Universidade Federal de Alagoas, 57072-900 Macei\'{o}, AL, Brazil
}%

\author{G. M. A. Almeida}
\email{gmaalmeida@fis.ufal.br}
\affiliation{%
 Instituto de F\'{i}sica, Universidade Federal de Alagoas, 57072-900 Macei\'{o}, AL, Brazil
}%

\begin{abstract}
The dynamics of two interacting quantum particles on a weakly disordered chain is investigated. 
Spatial quantum interference between them is characterized 
through the statistics of two-particle transition amplitudes, related to Hanbury Brown-Twiss correlations in optics. The fluctuation profile of the signal
can discern whether the interacting parties are
behaving like identical bosons, fermions, or distinguishable particles. 
An analog fully developed speckle regime displaying 
Rayleigh statistics is achieved for interacting bosons. 
Deviations toward long-tailed distributions echo
quantum correlations akin to  
non-interacting identical particles. In the limit of strong interaction, two-particle bound states obey 
generalized Rician distributions. 
\end{abstract}

\maketitle

Anderson localization is a universal phenomenon that underlies
wave physics \cite{evers08}. It is the outcome of a destructive interference of the waves due to a random potential.
In quantum mechanics, the problem is often addressed for noninteracting particles. The exponential growth 
in dimensionality makes it a daunting task to explore the onset of localization in an interacting multi-particle system. Interaction
can lead to involved physics such as many-body localization, which has recently seen significant progress \cite{abanin19}.

Yet, a system involving only two interacting particles delivers a rich set of features \cite{winkler06,bromberg09, lahini10,peruzzo10,krimer11,lahini12,corrielli13}. 
Indeed, many studies 
addressed the conditions in which the interaction 
can lead to an increase of the
localization length compared to the non-interacting case \cite{shepelyansky94,romer97,dias10,dufour12,lee14}.
Moreover, classical and quantum correlations have been explored in those systems even in the absence of interaction \cite{bromberg09, lahini10}. 
The quantum correlations in this case originate from the symmetrization of the wavefunctions to accomodate
the bosonic or fermionic character of the particles \cite{krimer11}. Quantum walks of two anyons was also investigated \cite{wang14,zhang22}. 
Other studies explored Bloch oscillations with a characteristic frequency doubling \cite{dias07, corrielli13} 
and dissipative two-particle dynamics \cite{rai15}, to name a few. 
In addition to displaying a variety of phenomena, two-particle systems enjoy a convenient
photonic implementation based on a square waveguide lattice using only classical sources of light \cite{longhi11-2,peruzzo10,krimer11,corrielli13,lee14}.

\begin{figure}[t!] 
\includegraphics[width=0.45\textwidth]{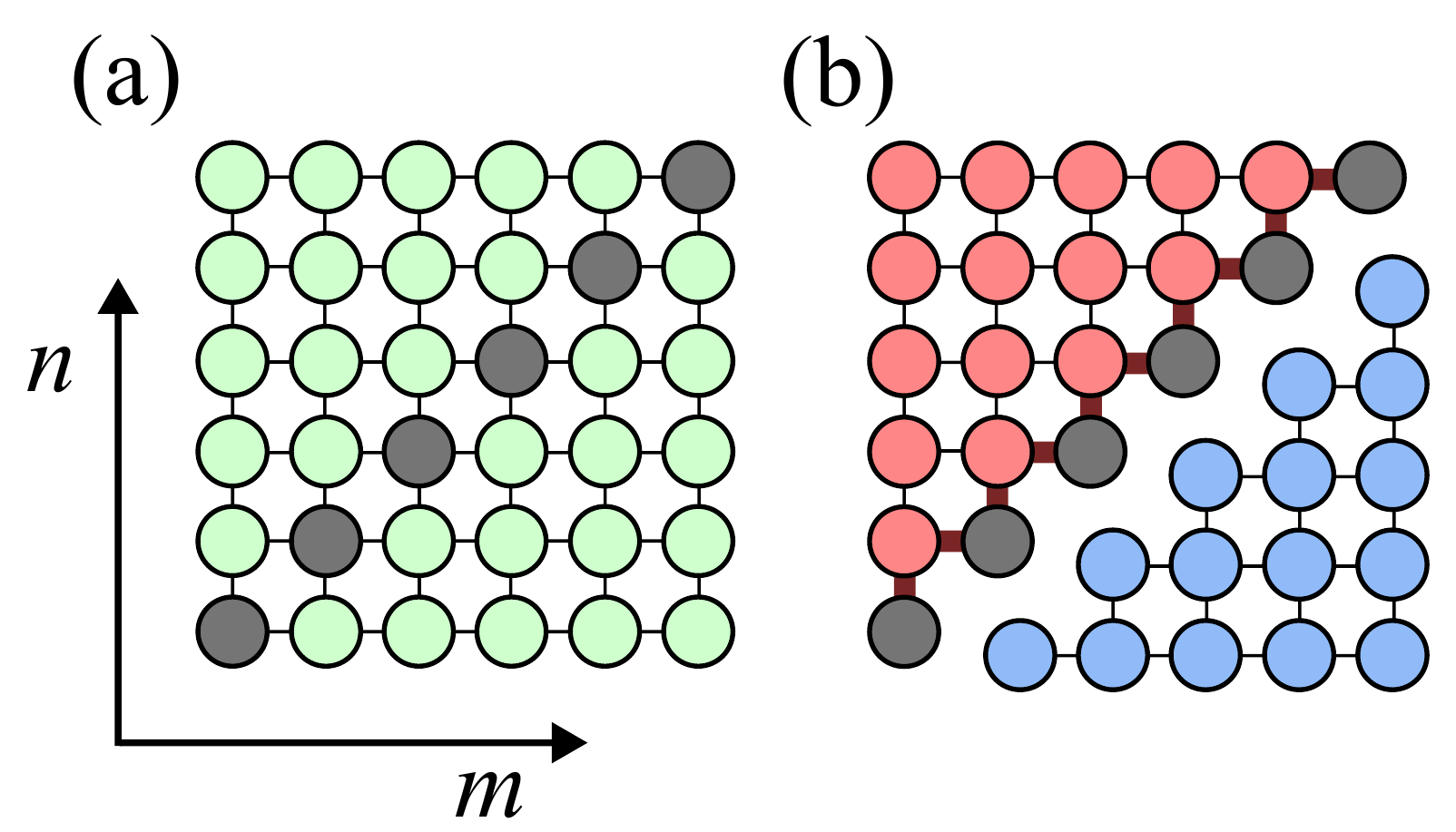}
\caption{\label{fig1} Two-particle Hamiltonian graph structure. (a) The state space of two distinguishable particles in 1D can
can be mapped onto a 2D array. Diagonal vertices represent states with double occupation (bound states). 
(b) By exploiting the symmetry with respect to the diagonal, the basis change $\ket{mn}^{\pm}=(\ket{mn}\pm\ket{nm})/\sqrt{2}$ decouple
the Hamiltonian into those describing identical bosons (red vertices) and spinless fermions (blue vertices). In the bosonic case, 
the coupling between bound states with the other vertices are renormalized by $\sqrt{2}$ (thick edges).
Considering a 2D photonic waveguide implementation,
each of these subspaces is achieved by setting the proper relative phase between two input beams at $(m,n)$ and $(n,m)$. Note that such a decoupling
is valid despite the strengths of disorder $W$ and interaction $U$. 
}
\end{figure}

The interplay between interaction and disorder in those systems is not trivial \cite{romer97} and depends on a number of factors, including the property that is being measured. Most characterizations require knowledge of many wavefunction amplitudes at a time (e.g. the participation ratio \cite{dias10,lee14}). 
In this letter we propose 
another route to obtain relevant information about the system. 
We are interested in the statistics of successive measurements of Hanbury Brown-Twiss type of correlations from a \textit{local} standpoint.  
In a coupled waveguide array implementation \cite{corrielli13} that means to monitor the beam intensity in single waveguide, accounting for joint probability of finding the particles at specific locations. What we get is a speckle pattern, to which we obtain
all the relevant density functions in detail. 
Surprisingly, the speckle contrast is able to 
precise the particle identity and the degree of interaction between them. It shows up as specific deviations from the Rayleigh/exponential fully-developed speckle regime.  


Tailored speckle generation finds a handful of applications \cite{bromberg14}.  
Our main goal here, though, is to 
explore the following question: what can a local speckle statistics tell us about the nature of the physical mechanisms involved in its generation?  
This statement is particularly appealing to rogue (freak) wave phenomena in optical and quantum systems. There has been a renewed interest
in the role of disorder on the generation of rare and short-lived wave amplitude spikes \cite{buarque22,buarque22-2}. In a recent work, Kirkby \textit{et al.} \cite{kirkby22} addressed Fock-space caustics in simple Bose-Hubbard models, which are also related to rogue events. Here we will realize that intrinsic quantum correlations due to particle identity lead to long-tailed distributions. Rogue waves are often studied as emergent phenomena in nonlinear Schrödinger equations \cite{dudley19} (that describe, for instance, Bose-Einstein condensates \cite{tan22}).
A bottom-up approach should therefore unveil key linear elements in driving anomalous fluctuations in quantum systems.  

%

\begin{figure}[t!] 
\includegraphics[width=0.35\textwidth]{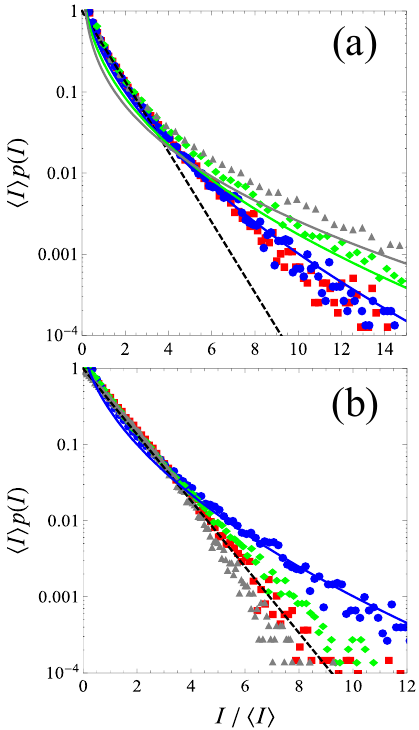}
\caption{\label{fig2} PDFs of local intensities $I$ evaluated at distinct times
for a single disordered array with $N=40$ sites with $W=0.01J$.
Statistics is taken from many operations of $e^{-iHt}$
up to $tJ=10^7$ in steps of size $100$. 
(a) Two non-interacting particles ($U=0$) are prepared at sites $(m,n)=(20,22)$ and
measured at $(p,q)=(23,26)$. The output for distinguishable particles is shown as green diamonds, with
the corresponding K-distribution function with shape parameter $\nu=1$ (green curve) with contrast $C\approx \sqrt{3}$.
When both particles are identical bosons or fermions with properly symmetrized input-output kets (red squares and blue circles)
we get another K-distribution with shape parameter $\nu=2$ (blue curve). 
Such a speckle profile features a contrast $C\approx \sqrt{2}$ as a consequence of entanglement due to wavefunction symmetrization.
Gray triangles 
stand for two distinguishable particles occupying the same location, with $m=n=20$ and $p=q=22$. 
In this case, only the bosonic subspace is involved. The correponding Weibull distribution is shown as the solid gray curve. 
The dashed black curve depicts the exponential density function (Rayleigh regime for $\sqrt{I}$). 
(b) In the presence of interaction ($U=1J$) the distribution for identical fermions remains the same but all the others get closer to an exponential form. The bosonic distribution 
fits right in but that corresponding to distinguishable particles display an extended tail. This is due to an interference
between Rayleigh-distributed and K-distributed phasors.
When only bound states are involved the tail retracts signalizing the onset of
modes with shorter localization length.
}
\end{figure}

Let us start by considering two interacting distinguishable particles (e.g. two electrons with opposite spins) in a linear chain with $N$ sites described by the Hamiltonian
\begin{align} \label{H2}
        H &= J\sum_{j=1}^{N-1}\left( a_{j+1}^{\dagger}a_{j} + b_{j+1}^{\dagger}b_{j} \right) \nonumber \\ 
& \,\,\,\,\,+ \sum_{j=1}^{N}\left[ \epsilon_{j}\left( a_{j}^{\dagger}a_{j} + b_{j}^{\dagger}b_{j} \right) + Ua_{j}^{\dagger}a_{j} b_{j}^{\dagger}b_{j} \right],
    \end{align}
 where $a_j$, $b_j$ ($a_j^{\dagger}$, $b_j^{\dagger}$) are the corresponding annihilation and creator operators at site $j$. $U$ is the local particle (respulsive) interaction strength, $J$ is the nearest-neighbor hopping constant, and $\epsilon_{j}$ is the onsite potential which we set randomly within the uniform interval $[-W/2,W/2]$, with $W$ being the disorder width. The Hilbert space is spanned by $N^2$ two-particle states $\ket{mn}=b_n^{\dagger}a_m^{\dagger}\ket{0}$, where $\ket{0}$ is the vacuum state.
 It is known that a basis change with respect to the ``diagonal'' double occupancy (bound) states $\ket{mn}^{\pm}=(\ket{mn}\pm\ket{nm})/\sqrt{2}$ ($m\neq n$) decouple the Hamiltonian in two parts  \cite{krimer11}. The symmetric combinations alongside bound states interact via a Bose-Hubbard Hamiltonian accounting for two identical bosons and the anti-symmetric part behave as non-interacting spinless fermions. Figure \ref{fig1} depicts the state-space structure. 
 Any local speckle pattern of the intensities will be therefore controlled by these two subspaces, each playing a distinct role. 

Before we elaborate the speckle formalism for the two-particle dynamics, it is appropriate to work out the transition amplitude statistics of a single-particle Anderson model. Consider the transition amplitude between sites $m$ and $p$ due to the time evolution operator $\mathcal{U}=e^{-iH^{(1)}t}$,
\begin{equation}
f_m^p = \langle p \vert \mathcal{U} \vert m \rangle = \sum_k a_{k}(m,p) e^{-iE_k t}= A e^{i\theta}.
\end{equation}
The phasor sum coefficients $a_{k}(m,p)=v_{k,m}v_{k,p}$ read from the local eigenfunctions $v_{k,j}=\langle j | E_{k} \rangle$ of the single-particle Hamiltonian $H^{(1)}$. 
When disorder is weak (large localization length) the set of $|a_k|$ is almost evenly distributed.
The phases $E_k t$ effectively behave as uncorrelated random variables uniformly distributed in $[0,2 \pi)$ at distinct times given $J \Delta t \gg 1$. If we 
get enough data in a time series, the real and imaginary parts of $f_m^p$ will obey
circular Gaussian statistics.
In turn, $A$ obeys the Rayleigh distribution
$p_{A}(A)=(A/\sigma^2)\exp{(-A^2/2 \sigma^2)}$, where $\sigma$ is a scale parameter, with the output phase $\theta$ being uniformly distributed \cite{goodmanbook}. 
The intensity $I = A^2$ then obeys the exponential distribution $p_I(I) = p_A(\sqrt{I})|dA/dI|= s^{-1}e^{-I/s}\equiv Exp(s)$,
with mean intensity $\langle I \rangle = s = 2\sigma^2$.
A relevant measure to discriminate between speckles is the ratio between the standard deviation of the intensity by its mean, namely the contrast $C$. A fully developed speckle obeying exponential statistics renders $C=1$. This gives us a reference to 
evaluate the degree of fluctuations of a given speckle pattern.  

Now that we have set up
the transition amplitude of a single particle as a random process, 
let us extended it to the case of two distinguishable particles when $U=0$.
Considering an input prepared at sites $(m,n)$, the transition to $(p,q)$ reads 
\begin{equation}\label{hmn}
h_{mn}^{pq} = 
f_{m}^{p}f_{n}^{q}=A_{1}A_{2}e^{i(\theta_1+\theta_2)}.
\end{equation}
The corresponding intensity is analogous to the two-particle correlation function $\langle a_{m}^{\dagger}b_{n}^{\dagger}b_n a_m  \rangle$, known
as Hanbury Brown-Twiss correlations in optics \cite{bromberg09, lahini10,lahini12,lee14}.
Each individual intensity in Eq. (\ref{hmn}) follows an exponential distribution, $I_i=A_i^2\sim Exp(s_{i})$. If we let 
$I_1$ and $I_2$ be independent random variables, it can be shown that their product $I=I_1 I_2$ obeys the K-distribution \footnote{Generalized K-distributions result from the product of two independent gamma distributions, which have
the exponential distribution as a particular case.}
\begin{equation}\label{kdis}
\mathcal{K}(I;\mu,\nu)=\frac{2\nu}{\mu \Gamma(\nu)}\left( \sqrt{\frac{I}{\mu}\nu} \right)^{\nu-1}K_{\nu-1}\left( 2 \sqrt{\frac{I}{\mu}\nu} \right),
\end{equation}
with shape parameter $\nu=1$. Therein $K_\nu(x)$ is a modified Bessel function of the second kind of order $\nu$ and $\mu=s_1s_2$ is the mean intensity. 
The contrast of a speckle obeying such a
of K-distribution is $C(\nu)=\sqrt{(\nu+2)/\nu}$. Hence, 
larger fluctuations are expected when two distinguishable particles are involved, that is $C=\sqrt{3}\approx 1.73$, even they are not interacting.
It is worth to highlight that $K$-distributions arise whenever some speckle intensity is known to obey exponential statistics but there is uncertainty about its mean $s$ \cite{hohmann10, goodmanbook}. 

When obtaining Eq. (\ref{kdis}) we assumed that $I_1$ and $I_2$ were independent. This is true for most input $(m,n)$
and output $(p,q)$ location pairs. However, some residual correlations can be present depending on where intensity measurements are being taken. This happens, for instance, when $|p-m|=|q-n|$ and disorder is weak. Both intensities become fully correlated ($I_1=I_2$) when the transition amplitude involve only bound states, i.e. $|h_{mm}^{pp}| = |f_{m}^{p}|^2 \sim Exp(s)$. The intensity $I=|f_{m}^{p}|^4$ then obeys a Weibull distribution $p_I(y) = \alpha^{-1}(2y)^{-1/2} e^{-\sqrt{2y}}$, where $y=I/\alpha$ and $\alpha = 2s^2$ is the mean intensity. The contrast now reads $C=\sqrt{5}\approx 2.24$.


So far we have seen that the intensity speckle statistics 
associated to non-interacting distinguishable particles
is typically long tailed. This stems from the
correlations present in the spectrum
of the two-particle Hamiltonian in Eq. (\ref{H2}). When $U=0$ its diagonal form reads $H=\sum_{k_1,k_2}^{N}E_{k_1k_2}b_{k_2}^{\dagger}a_{k_1}^{\dagger}\ket{0}\bra{0}a_{k_1}b_{k_2}$, with $\ket{0}$ being the vacuum and $E_{k_1k_2} = E_{k_1}+E_{k_2}$. As such the $N^2$ phases $E_{k_1k_2}t$
are combinations of two identical sets of $N$ single-particle inputs.
By analyzing it through the 2D mapping (Fig. \ref{fig1}), the observed non-Rayleigh statistics with higher contrasts is the result of structural correlations. We will see shortly how those correlations are partially destroyed when $U\neq 0$.

Let us now discuss the speckle profile of identical bosons and spinless fermions separately. 
In a photonic waveguide array, each set can be explored 
by injecting 
two coherent beams at locations $(m,n)$ and $(n,m)$ with the proper symmetric or antisymmetric phase relationship \cite{krimer11,lee14}. 
Given an input $\ket{\psi(0)} =(b_n^{\dagger}a_m^{\dagger}\pm b_m^{\dagger}a_n^{\dagger})\ket{0}$  
the transition amplitudes read $h_{mn(B)}^{pq}=(f_{m}^{p}f_{n}^{q} + f_{m}^{q}f_{n}^{p})\mathcal{N}$, with $\mathcal{N}=2^{-(\delta_{mn}+\delta_{pq})/2}$, for bosons and $h_{mn(F)}^{pq}=f_{m}^{p}f_{n}^{q} - f_{m}^{q}f_{n}^{p}$ for fermions. In both cases there is interference between K-distributed speckles.
This is expected since
we are now dealing with entangled input states. 
Indeed, such a quantum correlation manifests in the speckle statistics by delivering weaker fluctuations than those promoted by distinguishable particles.
To see this, consider (bound states excluded)    
$f_{m}^{p}f_{n}^{q} \pm f_{m}^{q}f_{n}^{p}=A_1e^{i\theta_1}+A_2e^{i\theta_2}$ is a two-component random phasor sum with independent K-distributed amplitudes $A_i\sim 2\sqrt{I_i}\mathcal{K}(\sqrt{I_i};\mu,1)$ with mean $\langle A_i \rangle=\pi \sqrt{\mu}/4$ and uniformly-distributed phases $\theta_i$. 
Note that we are assuming a common mean for both variables. This is a reasonable assumption 
for a weakly disordered chain. 
The output intensity speckle can be evaluated by means of a version of a modified Kluyver-Pearson formula \footnote{The Kluyver-Pearson formula gives the intensity distribution for a finite number of phasors of equal amplitude. If these amplitudes are random obeying some $p_{A}(A)$, one must evaluate the compound distribution $\int p(I|A)p_{A}(A)dA$, where $p(I|A)$ is the conditional intensity density function depending on knowledge of $A$ (see Sec. 3.2.4 of \cite{goodmanbook} for details).}. It results in another K-distribution [see Eq. (\ref{kdis})], now with shape parameter $\nu=2$ and mean $\mu'=2\mu$, that is $p_I(I)=\mathcal{K}(I;\mu',2)$.
The contrast is now $C=\sqrt{2}\approx 1.41$, lower than that obtained for distinguishable particles ($C=\sqrt{3}$). 
This is the
entanglement due to wavefunction symmetrization modifying the classical speckle signal. 
Figure \ref{fig2}(a) displays all the distributions obtained so far in agreement with the numerical simulations. 

\begin{figure}[t!] 
\includegraphics[width=0.45\textwidth]{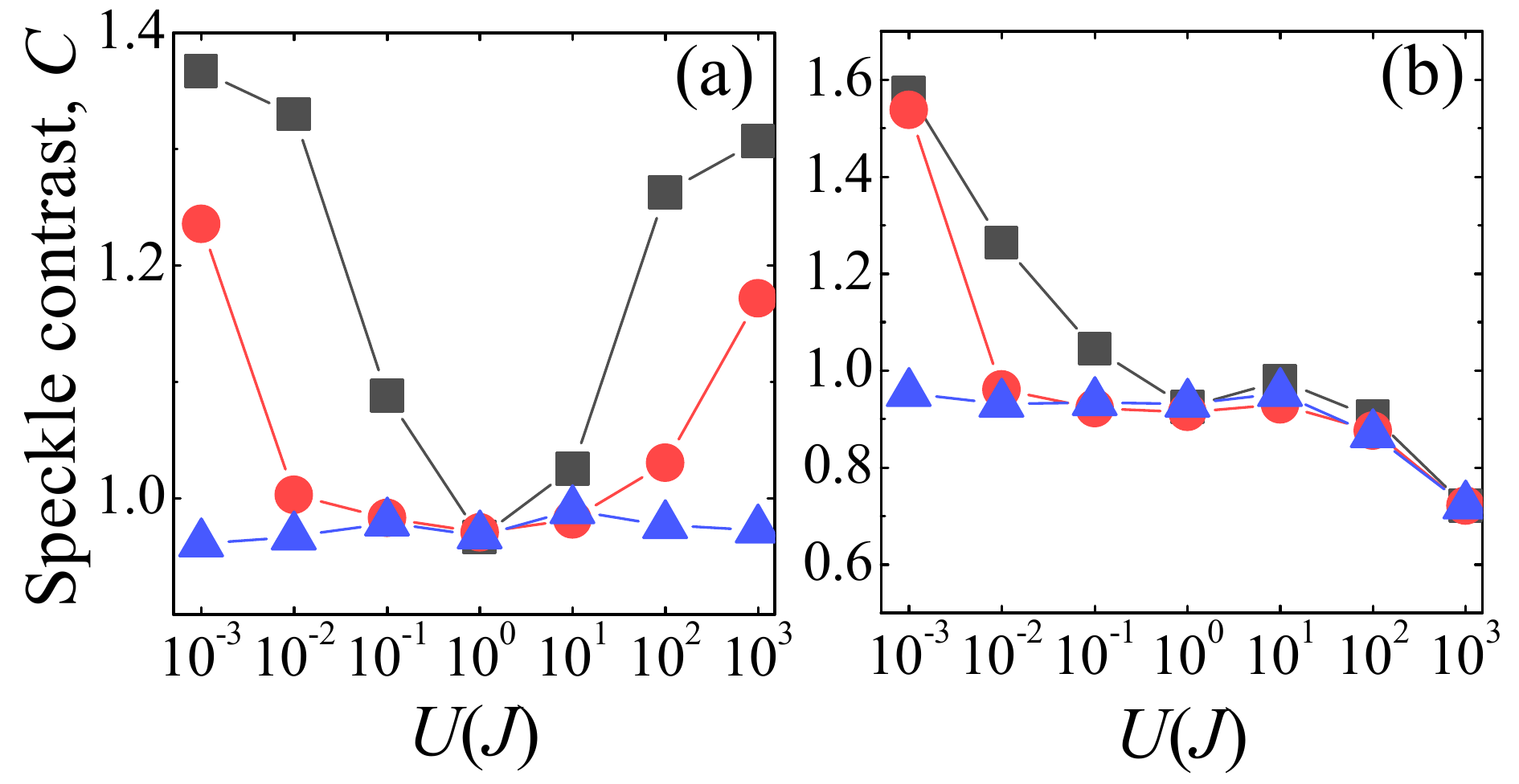}
\caption{\label{fig3} Local speckle contrast $C$ against interaction strength $U$ on a chain with $N=26$ sites and $W=0.01J$. 
For a given disorder sample, the statistics is taken within three distinct time windows, namely short ($tJ\in [0,\Delta]$; squares), intermediate ($tJ\in [10^{6},10^{6}+\Delta]$; circles), and long ($tJ\in [10^{9},10^{9}+\Delta]$; triangles), with $\Delta = 10^{5}$, in steps of $100$.
Contrast curves are averaged over 100 independent realizations of disorder. 
In (a) two bosons are loaded at sites $(10,11)$, with the intensity measurements being taken at $(13,16)$. Panel (b) depicts the case of bound states, with the two bosons placed at $(10,10)$ and measured at $(11,11)$. Contrast $C=1$ corresponds to exponential intensity statistics (fully developed speckle regime). 
}
\end{figure}

We are now ready to see how the presence of a local interaction 
between both particles 
modifies the speckle statistics.  
When $U\neq 0$ transition amplitudes between the quantum states can no longer be expressed in terms of single-particle wavefunctions. We expect that this 
symmetry loss in the two-particle spectrum may drive the intensity statistics toward the (fully developed) exponential regime. However, when we expand the dynamics in terms of the bosonic and fermionic subspaces (see Fig. \ref{fig1}), the latter is not affected by $U$. A given input $b_n^{\dagger}a_m^{\dagger}\ket{0}$ will evolve independently in each one of those subspaces, having symmetric and anti-symmetric components. 
The fermionic part maintains its K-distributed profile $\mathcal{K}(I;\mu',2)$. It thus suffices to examine the transition between bosonic states against $U$. 

Figure \ref{fig3} shows the contrast $C$ for several values of the interaction $U$ in distinct timescales.  When $U\ll J$, 
exponential statistics is only obtained in the long-time regime. That is, as the bosonic spectrum is slightly shifted it takes a while before the
random phasor sum underlying the
bosonic dynamics converges to a fully developed speckle [see Fig. \ref{fig2}(b)].  
When $U$ is weak, whether or not both bosons are loaded in the same site,
the short-time regime typically features higher fluctuations. Figure \ref{fig3}(a) [\ref{fig3}(b)]
indicates that these are reminiscent of the speckle pattern associated to the K-distribution (Weibull distribution).

At intermediate values of $U$, the particles are still free to access the whole bosonic subspace but the speckle signal 
associated to their symmetry that is present when $U=0$ is promptly lost.   
As $U$ increases, a smaller band of $N$ bound (B) states builds up apart from the scattering (S) part of the spectrum consisting
of $N(N-1)/2$ states \cite{winkler06,lahini12}. 
It is then convenient to 
express the transition amplitude as the phasor sum of the form $h_{mn(B)}^{pq} = \sum_{k\in S} b_k e^{i\phi_k}+\sum_{k'\in B} b_k' e^{i\phi_k'}$.
%
When $U\gg J$, if both bosons are injected in 
different sites they will display fermion-like correlations such as 
spatial anti-bunching \cite{lahini12} (the phasor sum running over the bound states becomes negligible). This is heralded as the high contrast seen in Fig. \ref{fig3}(a). 
It does not mean, however, that such a fermionic behavior will hold up at all times. Although transitions between scattered and bound states remain negligible, the speckle will eventually set as a fully developed one ($C\approx 1$) unless $U \rightarrow \infty$ (hard-core boson limit). 

We now realize that when loading two distinguishable particles in different locations $(m,n)$, the 
resulting speckle in $(p,q)$ comes as an interference between Rayleigh- and K-distributed phasors. Figure \ref{fig2}(b) shows that the speckle is nearly exponentially distributed aside from a pronounced tail (the contrast is numerically found to be $C\approx 1.05$). 
Here the fermionic correlations is holding the speckle from its full development. 

\begin{figure}[t!] 
\includegraphics[width=0.35\textwidth]{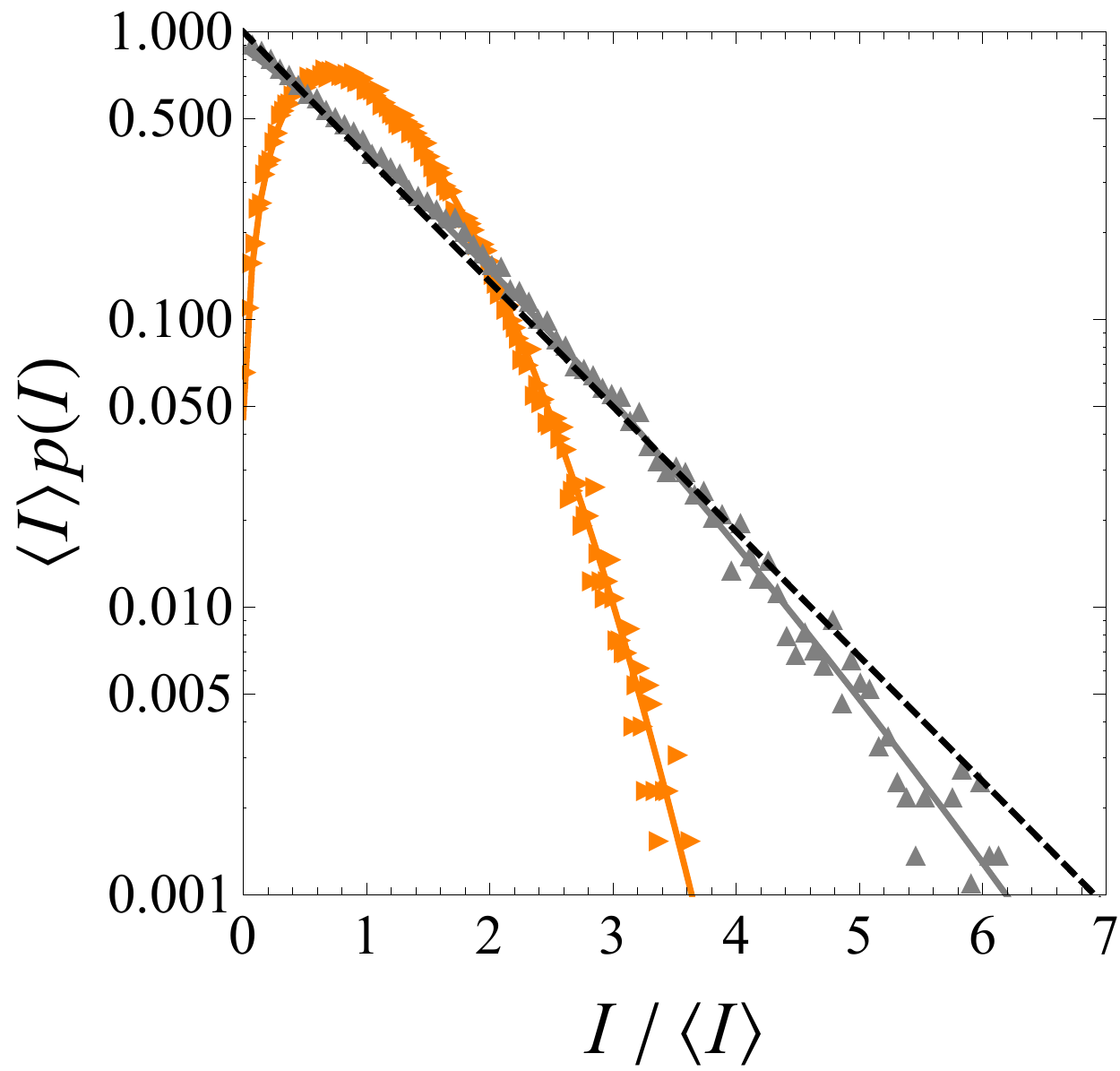}
\caption{\label{fig4} 
Distributions of the intensities associated to bosonic bound state transitions. Here, $I=|h_{mm}^{pp}|^2$, with $m=20$ and $p=22$,  
considering a chain with $N=40$ sites in the strong $U$ regime. Statistics is taken
on a single disorder sample with $W=0.01J$ evolving up to time $tJ=10^7$ in steps of 100.   
Up (right) Triangles represent the case for $U=200J$ ($U=500J$). 
Solid curves are the compound Rician fittings obtained by isolating the four greatest amplitudes of the corresponding random phasor sum to build $g(r)$ (see text).
The integral $\int R(I|r) g(r)dr$ is evaluated numerically. Note that the necessary number of phasors involved to make $g(r)$ depends on both $U$ and 
the distance between input and output locations. When the distribution is similar to an exponential one, aside from a slightly retracted tail, a standard Rician distribution with only one dominant phasor
(when $g(r)$ is a Dirac delta function) should be enough.
The exponential distribution is displayed as a dashed curve for reference.
}
\end{figure}

A noteworthy trend that occurs in the regime of strong interaction $U$ is the decrease in contrast when only bound states are involved [see Fig. \ref{fig3}(b)].
Early signs of this behavior can actually be noticed for intermediate $U$ [see the tail retraction of the associated distribution in Fig. \ref{fig2}(b)].
That can be readily explained in terms of the energy pulling effect taking place in the band of bound states \cite{winkler06,lahini12}. Even if the disorder $W \ll J$, the effective hopping strength within the band will eventually diminish to the point that strongly-localized bound states prevail. In a sum involving random phasors, it fosters asymmetry between their amplitudes. One or a few phasors will stand out compared to the remaining terms that amount to an exponentially-ditributed intensity with mean $s_n$. In the simplest case of only one dominant phasor with intensity $I_0$, we get the Rician distribution \cite{goodmanbook} 
\begin{equation}
R(I;r) =s_n^{-1}e^{-(r+I/s_n)} \mathcal{I}_{0}(2\sqrt{I r/s_n}),
\end{equation}
where $\mathcal{I}_{0}(x)$ the modified Bessel function of the first kind of order zero and $r=I_0/s_{n}$.
The contrast in this case is $C(r)=\sqrt{1+2r}/(1+r)$.
If more than one dominant phasors is set apart
we can compound the distribution above over different $r$ to obtain $p_I(I)=\int R(I|r) g(r)dr$
where $R(I|r)$ is the Rician distribution conditioned on knowledge of $r$ and $g(r)$ is its probability density function. 


The above generalized Rician distributions will also display a lower contrast compared to a fully developed speckle. It ultimately depends on $U$ as well as the distance between
input and output sites given the typical spatial profile of localized modes.
That is, if such a distance is large enough we expect
$g(r)\sim Exp(s_0)$ and then the Rician compound becomes an exponential distribution with mean $\langle I \rangle = s_n+s_0$.
It is valid to mention that there is a non-monotonic relationship between $U$
and the degree of Anderson localization disordered 
two-particle systems \cite{dias10}.
To see the generalized Rician distributions in activity here, let us turn our attention to the regime of strong $U$. Figure \ref{fig4} confirms the predicted statistics for transitions involving bound states. 


We have seen that fluctuations associated to local intensity measurements can 
disclose subtle quantum correlations. 
Non-Rayleigh speckles can be 
extracted from the time evolution of two quantum particles. 
It ranges from low contrast forms obeying generalized Rician distributions
to K-distributed speckles that display higher-than-exponential fluctuations.
The two-particle dynamics can be promptly adapted to a  square photonic waveguide array loaded with classical light \cite{longhi11-2,peruzzo10,krimer11,corrielli13,lee14}. 
The different speckle patterns is then 
obtained upon setting the desired input phase relationship (so as to activate bosonic and/or fermionic behavior) and controlling the
detuning between the diagonal waveguides
and the others \cite{longhi11-2}.

Besides having immediate applications is optics, our results 
apply to the characterization of quantum systems in general. 
We have shown that 
local measurements in the computational basis is able to capture subtle quantum correlations involving identical particles. 
Even when both particles are distinguishable their resulting speckle
distribution display a contrast $C>1$, a property
that can be traced back to fermionic correlations. 
The speckle corresponding to two bosonic particles also feature a higher contrast in both weak and strong $U$ limits. It takes some time before their intrinsic correlations are washed out and we get a fully developed speckle. This is an interesting feature that allows one to manipulate the speckle statistics while maintaining the overall dynamical pattern \cite{bromberg14}. %
This is yet another manifestation of  entanglement due to particle identity 
that meets practical applications \cite{morris20}.
Future works may delve into the speckle response in the multi-particle level \cite{cai21,giri22} and its relationship with other forms of entanglement.    

This work was supported by CNPq, CAPES (Brazilian agencies),
and FAPEAL (Alagoas state agency). 


%

\end{document}